\documentstyle[12pt,psfig]{article}

\newcommand{\lbl}[1]{\label{eq:#1}}
\newcommand{\rf}[1]{(\ref{eq:#1})}
\newcommand{\vs}[1]{\rule[- #1 mm]{0mm}{#1 mm}}
\newcommand{\be}{\vs{2}\begin{equation}}
\newcommand{\eq}{\vs{2}\begin{equation}}
\newcommand{\en}{\\[2mm]\end{equation}}
\newcommand{\bea}{\begin{eqnarray}}
\newcommand{\ena}{\end{eqnarray}}

\newcommand{\NP}[1]{Nucl.\ Phys.\ {\bf #1}}

\newcommand{\AN}[1]{Ann. Phys. {\bf #1}}

\newcommand{\PR}[1]{Phys.\ Rep.\ {\bf #1}}

\newcommand{\ZP}[1]{Z.\ Phys.\ {\bf #1}}

\begin{document}

\input{epsf}

\begin{titlepage}

\renewcommand{\thefootnote}{\fnsymbol{footnote}}

\today

\rightline{IPNO-DR 99 23}
\rightline{CPT-99/P.3915}

\vspace*{1cm}

\begin{center}

\indent

{\Large{\bf Contributions of order ${\cal O}(m_{\rm quark}^2)$  to
$K_{\ell 3}$ form factors and unitarity of the CKM matrix}}
\footnote{Work supported in part by TMR, EC-Contract No. ERBFMRX-CT980169
(EURODA$\Phi$NE)}

\vspace{1.5cm}

{\bf{N.H. Fuchs}} 

\indent
{\sl Physics Department,
Purdue University\\
West Lafayette IN 47907, USA}\\[0.5cm]

{\bf{M. Knecht}} 

{\sl Centre  de Physique Th{\'e}orique,
       CNRS-Luminy, Case 907\\
    F-13288 Marseille Cedex 9, France\\[0.5cm]}

{\bf{J. Stern}} 

{\sl     Groupe de Physique Th\'eorique,
        Institut de Physique Nucl\'eaire\\
        F-91406 Orsay Cedex, France }

\vspace{2cm}

\end{center}

\begin{abstract}

The form factors for the $K_{\ell 3}$ semileptonic decay are computed to 
order $O(p^4)$ in generalized chiral perturbation theory.  The main 
difference with the standard $O(p^4)$ expressions consists in  contributions 
quadratic in quark masses, which are described by a single divergence-free
low-energy constant, $A_3$.  A new simultaneous analysis is presented
for the CKM matrix element $V_{us}$, the ratio $F_K/F_{\pi}$, $K_{\ell 
3}$ decay rates and the scalar form factor slope $\lambda_0$.
This framework easily accommodates the precise value for $V_{ud}$
deduced from superallowed nuclear $\beta$-decays.

\end{abstract}

\vspace{5cm}

\end{titlepage}

\begin{titlepage}

$ $

\end{titlepage}

\indent

\pagestyle{plain}  
\setcounter{footnote}{0}                                                      
\setcounter{equation}{0}
\setcounter{subsection}{0}
\setcounter{table}{0}
\setcounter{figure}{0}

\renewcommand{\theequation}{\arabic{section}.\arabic{equation}}
\section{\bf Introduction}

Together with low-energy $\pi$--$\pi$ scattering \cite{GL84b,GL84c}, 
the semi-leptonic
$K_{\ell 3}$ \cite{GL85b} and $K_{e4}$ \cite{bijnenskl4,rigg} decays have 
been among 
the first applications of standard chiral perturbation theory (S$\chi$PT) 
\cite{GL84a,GL84c,GL85a} to be studied at the one-loop level.
At this order, the mesonic form factors that describe these decays do not
contain the poorly known low energy constants $L_8$
and $L_6$, and consequently they may be expected to be less sensitive
to the size of the chiral condensate $\langle\bar q q\rangle$ than,
e.g., the $\pi-\pi$ s-wave \cite{FSS,KMSF}. 
In particular, the $K_{\ell 3}$ form
factors at ${\cal O}({\rm p}^4)$ merely involve, besides the masses and decay 
constants of the pseudoscalar states, the low energy constant $L_9$, whose 
value can be obtained \cite{GL85b} from the experimentally measured 
 charge radius of the pion \cite{pirad}.
This fortunate circumstance has been used in the
past to extract the CKM matrix element $V_{us}$ from the $K_{e3}$ decay
rates \cite{lroos}, and presently this extraction is still considered by 
the Particle 
Data Group (PDG) compilation \cite{PDG98} to remain
the most accurate and least model dependent. Yet this determination of
$V_{us}$ relies on a model-dependent estimate of ${\cal O}(m_{\rm quark}^2)$ 
contributions to the form factor $f_+(0)$ (the notation will be given below), 
which, in S$\chi$PT, arise as contributions of order ${\cal O}({\rm p}^6)$.
With the new careful and accurate determinations of $V_{ud}$  from
superallowed nuclear $\beta$-decays \cite{Ha90,ENAM95,WEIN98}, the size of 
these corrections 
is required to be comparable to the (parameter-free) genuine 
${\cal O}({\rm p}^4)$ contribution in order to preserve the unitarity of the 
CKM matrix. It is then legitimate to ask whether similarly
important ${\cal O}(m_{\rm quark}^2)$ contributions would not affect the
parameter-free S$\chi$PT  prediction for the slope $\lambda_0$ 
of the $K_{\mu 3}$ scalar form factor at order ${\cal O}({\rm p}^4)$.
 In principle, these questions can be answered by performing 
the full two-loop S$\chi$PT calculation of the $K_{\ell 3}$ form factors,
provided the several new ${\cal O}({\rm p}^6)$ counterterms that will 
contribute can be measured independently or estimated in a reliable 
way~\footnote{The structure of the order ${\cal O}({\rm p}^6)$ effective 
lagrangian of S$\chi$PT has been discussed in Refs. \cite{FeaSch,BCE1999}.}. 
The present status of this 
enterprise is limited~\footnote{Unfortunately, the ${\cal O}({\rm p}^6)$ 
calculation of Ref. \cite{Post} only considers a very specific 
combination of the {\it slopes} of the 
$f_+^{K\pi}$ and of the mesonic electromagnetic form factors.} to the 
evaluation of the so-called chiral double-logs 
\cite{BCE1998}, which, although only part of the full two-loop corrections, 
do not seem to point towards huge effects coming from the chiral loops 
themselves (see in particular the numerical results in Table 1 of 
Ref.~\cite{BCE1998} and the comments preceding it). 

The purpose of the present work is to address some of these issues from the 
point of view of generalized chiral perturbation theory (G$\chi$PT) 
\cite{FSS,SSF,KS}.
The expressions for the form factors of semileptonic decay of kaons  to  
${\cal O}({\rm p}^4)$  in G$\chi$PT have never been published (a discussion 
of the $K_{\ell 4}$ form factors at order ${\cal O}({\rm p}^2)$ in G$\chi$PT 
can be found in Ref.~\cite{Kl4G}), although they have existed for some
time and have been partially reported on various occasions. Keeping in
mind the well-known and important exception of the phase of the $K_{e4}^+$
decay amplitude, the $K_{\ell 3}$ and $K_{e4}$ form factors are
indeed found to be to a large extent independent of the size of the
condensate $\langle\bar q q\rangle$. 
The main consequence of the
modified chiral counting ($m_q \sim {\cal O}({\rm p}))$ is that in G$\chi$PT
the form factors receive a contribution quadratic in quark masses
already at order ${\cal O}({\rm p}^4)$, in addition to the standard 
${\cal O}({\rm p}^4)$ expressions. Furthermore, these new 
${\cal O}(m_{\rm quark}^2)$
contributions, which within S$\chi$PT would count as ${\cal O}({\rm p}^6)$,
{\bf are all related}. They all stem from a single term of the
${\cal L}_{(2,2)}$ component of the effective Lagrangian 
${\cal L}_{{\rm eff}}$ and they are
described by a single divergence-free low-energy constant $A_3$ (see
Ref.~\cite{KMSF} and Appendix A for notation). (This statement is exact in
the case of $K_{\ell 3}$ form factors $f_+$ and $f_-$, whereas in the
case of $K_{e4}$ it remains true for the dominant ${\cal O}(m_s^2)$ terms.)
It thus appears that the ${\cal O}({\rm p}^4)$ G$\chi$PT offers a predictive
description of the terms quadratic in quark masses which is
non-trivial compared to the one-loop S$\chi$PT (no quadratic terms)
and yet much simpler than the standard ${\cal O}({\rm p}^6)$ order, in which
other unknown constants should contribute in addition to the
${\cal O}({\rm p}^4)$ G$\chi$PT terms driven by the constant $A_3$.
This framework, which is as systematic as the standard expansion, suggests 
a new simultaneous analysis of $V_{us}$,
$K_{\ell 3}$ decay rates, $F_K/F_{\pi}$, and of the scalar slope
$\lambda_0$ which may be of interest in connection with the constraint
of unitarity of the CKM matrix and with the forthcoming new 
$K_{\ell 3}$ data. A closely related application, namely the determination 
of the ${\cal O}({\rm p}^4)$ constant $L_3$ from the $K_{e4}^0$ decay rate, 
will be presented elsewhere~\cite{FKSunpub}, together with a full discussion 
of the $K_{\ell 4}$ form factors at order ${\cal O}({\rm p}^4)$ in 
G$\chi$PT. 

The present stage of our analysis involves one additional limitation, to the 
extent that no electromagnetic corrections are included, unless
explicitly stated. For this reason we postpone a detailed analysis of
the isospin asymmetry in the $K_{e3}^0$ and $K_{e3}^+$ decay rates 
that is due to the mass difference $m_d-m_u$. We just check that this
asymmetry is consistent with the G$\chi$PT treatment of $\pi^0-\eta$
mixing within errors.

The paper is organized as follows: Section 2 provides the necessary 
expressions of the $K_{\ell 3}$ form factors and of the $\pi^0-\eta$ mixing 
angles in G$\chi$PT. Implications for the determinations of $V_{us}$ and of 
the ratio $F_K/F_\pi$ of pseudoscalar decay constants are discussed in 
Section 3. The slopes of the $K_{\ell 3}$ form factors are considered in 
Section 4. Concluding remarks are presented in  Section 5. Details on the 
${\cal O}({\rm p}^4)$ structure of the G$\chi$PT effective Lagrangian and on 
its renormalization are presented in Appendix A. Useful expressions for the 
pseudoscalar masses and decay constants have been gathered in Appendix B.

\addtocounter{section}{0}
\setcounter{equation}{0}
\section{\bf $K_{\ell 3}$ form factors in G$\chi$PT to ${\cal O}({\rm p}^4)$}

We consider the two semileptonic decay channels
\bea
 K^+(p) &\rightarrow& \pi^0 (p') \ell^+ (p_\ell) \nu _\ell (p_\nu) \hspace{1cm}
[K_{\ell 3}^+] \nonumber \\
K^0(p) &\rightarrow &\pi^- (p') \ell^+ (p_\ell) \nu_\ell (p_\nu) \hspace{0.9cm}
[K_{\ell 3}^0]
\label{channels}.
\ena
The symbol $\ell$ stands for $\mu$ or $e$. As stated before, we do not consider
electromagnetic corrections.   The processes (\ref{channels})
are then described by four form factors, $f^{K^+\pi^0}_\pm
(t)$ and $f^{K^0 \pi^-}_\pm (t)$, which depend on $t = (p'-p)^2 = (p_l +
p_\nu)^2$, the square of the four momentum transfer to the leptons, and which 
are defined in terms of the hadronic matrix elements of the charged 
strangeness changing QCD vector current as follows:
\bea
<\pi^0(p')\vert ({\overline s}\gamma_\mu u)(0)\vert K^+(p)>
&=& \frac{1}{\sqrt{2}}\big[(p+p')_\mu f^{K^+\pi^0}_+ 
+ (p-p')_\mu f^{K^+\pi^0}_-\big]
\nonumber\\
<\pi^-(p')\vert ({\overline s}\gamma_\mu u)(0)\vert K^0(p)>
&=& (p+p')_\mu f^{K^0\pi^-}_+ + (p-p')_\mu f^{K^0\pi^-}_- .
\ena

\subsection{G$\chi$PT expressions of the $K_{\ell 3}$ form factors}

The one-loop G$\chi$PT expressions (using the notation of reference
\cite{GL85b}) for the form factors are summarized below. We start with the 
two form factors $f^{K^0 \pi^-}_+ (t)$ and $f^{K^+\pi^0}_+ (t)$, which are in 
practice sufficient for the description of the electron decay modes 
$K_{e 3}^+$ and $K_{e 3}^0$, and keeping 
isospin breaking contributions due to the quark mass difference $m_d - m_u$.
For the $K_{\ell 3}^0$ channel, which is somewhat simpler, since  
$\pi^0-\eta$ mixing only enters the loop contributions, we obtain
\bea \label{fpK0}
f_{+}^{K^0\pi^-}(t) &=& 1 
\ +\  H_{\pi^+K^0}(t)\ +\ {1\over 2}\,H_{\pi^0K^+}(t)\ +
\ {3\over 2}\,H_{\eta K^+}(t)\nonumber\\
&+& \sqrt{3}\,\stackrel{o}{\varepsilon}\,\bigg[ H_{\pi K}(t)\ -\ H_{\eta K}(t)
\bigg] \nonumber \\
&+& \bigg[ {1\over 8}{\widehat m}^2\,\xi^2\,+\,{1\over 2}{\widehat
m}^2A_3\bigg]\,(r-1)^2\,\big(1+{1\over R}\big) \;,
\ena
with (the definitions of the loop functions $M_{PQ}^{\rm r}\,(t)$, $K_{PQ}(t)$ and 
$L_{PQ}\,(t)$ that we use below can be found in Ref.~\cite{GL85a})
\be
H_{PQ}(t)\ =\ {1\over{F_\pi^2}}\,\bigg[t\,M_{PQ}^{\rm r}\,(t)\ -\ L_{PQ}\,(t)\ +
\ {2\over 3}\,L_9^{\rm r}\,t\,\bigg]\ .
\en
Here $\stackrel{o}{\varepsilon}$ denotes the leading order $\pi
^0-\eta$ mixing angle,
\be
\stackrel{o}{\varepsilon}\ =\  {\sqrt{3} \over 4 R} \bigg[ 1 -
{\Delta_{GMO} \over M_{\eta}^2 - M_{\pi}^2} + {2 \over 3}\, (r_2 - r)\ 
{r - 1 \over r+1}\  {M_{\pi}^2 \over  M_{\eta}^2 - M_{\pi}^2} \bigg] \;,
\label{eps0}
\en
where 
\bea
r&\equiv&m_s/{\widehat m} \nonumber \\
r_2 &\equiv& 2 {M_K^2 \over M_{\pi}^2} - 1  \\
\Delta_{GMO} &\equiv& 3 M_{\eta}^2 - 4 M_K^2 + M_{\pi}^2 \nonumber \;.
\ena
We work only at first order in the quark mass difference $(m_d -
m_u)$, {\it i.e.} we consider only terms that are at most of order
${\cal O}(1/R)$ \footnote{We have however kept the pion mass difference, 
which is mainly an electromagnetic effect~\cite{GLPR,Dasetal}, in the loop 
contributions.}, where 
\be
R = {m_s - {\widehat m} \over m_d - m_u}.
\en
The lowest order S$\chi$PT value for the $\pi^0-\eta$ mixing angle, 
$\stackrel{o}{\varepsilon}_{\rm st} = {\sqrt{3}
\over 4 R}$, is recovered by dropping, in Eq.~(\ref{eps0}), the last two 
terms, which are counted as
order ${\cal O}({\rm p}^4)$ in S$\chi$PT. The $\xi^2$ and $A_3$ terms in
Eq.~(\ref{fpK0}) are ${\cal O}({\rm p}^4)$ contributions in G$\chi$PT (see the
detailed formulas for the effective Lagrangian in Appendix \ref{knecht}), but 
are absent at this order in S$\chi$PT.

For the $K_{\ell 3}^+$ decay mode, we find
\bea \label{fKpfK0}
f_{+}^{K^+\pi^0}(t) &=& f_{+}^{K^0\pi^-}(t) \times \left\{ 1\ +\
{{\sqrt{3}}\over 2}( \varepsilon_1\ +\ \varepsilon_2 ) \right.\\
&& \qquad  - \left.
\ {1\over\sqrt{3}}{\widehat m}^2\xi^2(r-1)^2\,\stackrel{o}{\varepsilon}\ -
\ {\widehat m}^2 C_2^P(r-1)^2\,\bigg[ {1\over R}\ +\
{{2\stackrel{o}{\varepsilon}}\over{\sqrt{3}}}\bigg] \right\}\;,
\nonumber  
\ena
where the $\pi^0-\eta$ mixing angles at order ${\cal O}({\rm p}^4)$,
$ \varepsilon_1$, $\varepsilon_2$, are defined
in section \ref{epsilons} below. Although the constant $C_2^P$ corresponds to 
a counterterm of ${\cal L}_{(2,2)}$ (see Eq.~(\ref{L22}) in Appendix 
\ref{knecht}) that violates the Zweig rule, it is not expected 
to be suppressed, since this violation occurs in the $0^-$ channel. 

At zero momentum transfer, Eq.~(\ref{fpK0}) gives
\bea \label{fplus0}
f_{+}^{K^0\pi^-}(0) &=& 1 
\ -\ {1\over{256\pi^2F_\pi^2}}\,\bigg\{ 2\,(M_{\pi^+}^2+M_{K^0}^2)\,h_0\big(
{{M_{\pi^+}^2}\over{M_{K^0}^2}}\big)\nonumber\\
&& + \quad  (M_{\pi^0}^2+M_{K^+}^2)\,h_0\big(
{{M_{\pi^0}^2}\over{M_{K^+}^2}}\big) +
3\,(M_{K^+}^2+M_{\eta}^2)\,h_0\big(
{{M_{K^+}^2}\over{M_{\eta}^2}}\big)\\ 
&& + \quad  2 \, \sqrt{3}\stackrel{o}{\varepsilon}
\,\bigg[(M_{\pi}^2+M_K^2)\,h_0\big(
{{M_{\pi}^2}\over{M_{K}^2}}\big)\,-\,(M_{K}^2+M_{\eta}^2)\,h_0\big(
{{M_{K}^2}\over{M_{\eta}^2}}\big)\bigg] \bigg\} \nonumber \\
&+& \bigg[ {1\over 8}{\widehat m}^2\xi^2 + {1\over 2}{\widehat
m}^2A_3\bigg]\,(r-1)^2\, \left(1+{1\over R}\right)\nonumber \  ,
\ena
with
\be
h_0(x)\,=\,1\,+\,{{2x}\over{1-x^2}}\,\ln x\ .
\en
To recover the expression at order ${\cal O}({\rm p}^4)$ in
S$\chi$PT \cite{lroos,GL85b} for $f_{+}^{K^0\pi^-}(0)$, one replaces 
$\stackrel{o}{\varepsilon}$ by the leading order S$\chi$PT value 
$\stackrel{o}{\varepsilon}_{\rm st}$, and one simply drops the 
last line in Eq.~(\ref{fplus0}). Notice that this last contribution is the 
only correction of order ${\cal O}({\rm p}^4)$ that does not vanish in the 
large-$N_c$ limit of QCD. The fact that the ${\cal O}({\rm p}^4)$ corrections 
to $f_{+}^{K^0\pi^-}(0)$ in S$\chi$PT vanish altogether in the large-$N_c$ 
limit might provide a natural explanation why contributions of the 
${\cal O}({\rm p}^6)$ counterterms could be comparatively sizeable.

The expressions of the form factors $f_{-}^{K^0\pi^-}(t)$ and 
$f_{-}^{K^+\pi^0}(t)$ are rather cumbersome if ${\cal O}(m_d-m_u)$ effects 
are included. Since we do not need the latter in this case, we give only the 
common expression of these two form factors in the isospin limit, which is
\bea
f_-^{K\pi}(t) &=& {1\over 2}{\widehat m}\xi (r-1)\ -\ {1\over 4}
{\widehat m}^2\xi^2 (r-1)(r+3)\ -
\ {\widehat m}^2\xi{\tilde\xi}(r-1)(r+2)\nonumber\\
&+& {1\over 2}{\widehat m}^2(A_1+B_1)(r^2-1)\ +
\ {1\over 2}{\widehat m}^2(A_2-2B_2)(r-1)\nonumber\\
&+& {\widehat m}^2D^S(r-1)(r+2)\ +\ {1\over 4}[ 5\mu_{\pi}-2\mu_K-3\mu_{\eta}]
\\
&-& {1\over{4F_0^2}}\,\big[
5\,t - 2\,M_{\pi}^2 - 2\,M_K^2 + 16{\widehat
  m}^2(A_0 + 2\,Z_0^S)(r+1)\big] \,K_{\pi K}(t)\nonumber\\ 
&-& {1\over{4F_0^2}}\,\big[
3t-2M_{\pi}^2-2M_K^2 + 4{\widehat m}^2(A_0+2Z_0^P)(r-1)(r+3)\big]
\,K_{K\eta}(t) \nonumber\\
&-&{3\over{2 F_0^2}}\,(M_K^2 - M_{\pi}^2)\left\{
\left[ M_{K\pi}^{\rm r}\,(t)\,+\,{2\over 3}\,L_9^{\rm r}\right]\ +
\ \left[ M_{K\eta}^{\rm r}\,(t)\,+\,{2\over 3}\,L_9^{\rm r}
\right]\right\}\nonumber \;.
\ena
For the ease of comparison, we may rewrite this G$\chi$PT expression for 
the form factor $f_-(t)$ in terms of the corresponding S$\chi$PT expression: 
\bea 
f_-^{K\pi;std}(t) &=& \frac{F_K}{F_{\pi}}- 1  - \frac{2\,L_9^r\,
      \left( M_K^2 - M_{\pi}^2 \right)} {F_\pi^2}\nonumber \\
&&+  \, {\frac{\left(2\,M_K^2 + 2\,M_{\pi}^2 - 3\,t
      \right)}{4\,F_\pi^2}} \, K_{K \eta}(t) \nonumber \\
&&+ \,  {\frac{\left(2\,M_K^2 + 2\,M_{\pi}^2 - 5\,t \right)
      }{4\,F_\pi^2}} \,K_{\pi K}(t) \nonumber \\
&&-\, \frac{3\,\left( M_K^2 - M_{\pi}^2 \right) \,
      }{2\,F_\pi^2} \,\left[{M^r}_{K\eta}(t) + {M^r}_{K \pi}(t)\right]\;, \\   
&& \nonumber \\
f_-^{K\pi}(t) &=& f_-^{K\pi;std}(t) \nonumber \\ \label{fminusgen}
&&- \,\frac{1}{4}\left(\frac{F_K} {F_{\pi}} - 1 \right)^2\,
        \left({\frac{F_K^2}
            {{F_{\pi}}^2}} + 2\frac{\,F_K}{F_{\pi}} - 1 
          \right)  \nonumber \\
&&-   \,\frac{1}{2}\,{m^2}\,{A_3}\,(r-1)(r+3)    \nonumber \\
&&- \,{\frac{{\Delta_{GMO}}\,\left( r + 3 \right) }{4\,{F_\pi^2}\,
      \left( r-1 \right) }} \, K_{K\eta}(t) - 
  \frac{M_{\pi}^2\,\left( 1 + r \right) \,
      \hat{\epsilon}(r)}{F_\pi^2}\,K_{\pi K}(t)\;,
\ena
where 
\bea
\hat{\epsilon}(r) &\equiv& {4 \, {\widehat m}^2 \, (A_0 + 2\,Z_0^S) \over
M_{\pi}^2} = 2 {r_2 - r \over r^2 - 1} (1 + 2 \zeta) \nonumber \\
\zeta &=& Z_0^S/A_0 \;,
\ena
and the ${\cal O}({\rm p}^4)$ expression
\bea \label{mxiOp4}
\hat{m}\xi &= &{1 \over r-1} \left( {{F_K^2} \over {F_{\pi}^2}}
- 1 \right) \nonumber \\
&& - (r+1) \, \hat{m}^2 \, (A_1 + B_1) - \hat{m}^2 \, (A_2 - 2 B_2) -
(r+3) \, \hat{m}^2 \, A_3 \nonumber \\
&& - 2 \,(r+2) \, \hat{m}^2 D^S  \\
&& + 2 \, \hat{m}^2 \, \xi^2 + 2 \, (r+2) \, \hat{m}^2 \xi \tilde{\xi}
\nonumber  \\ 
&& - {1 \over 2(r-1) } \, (5 \mu_{\pi} - 2 \mu_K - 3 \mu_{\eta}) \nonumber 
\ena
has been used.  It is remarkable that, in Eq.~(\ref{fminusgen}), {\em
all} ${\cal L}_{(2,2)}$ constants -- {\em except} for $A_3$ -- have
been absorbed into the renormalization of the decay constants. Notice also 
that, in contrast to $f_+(t)$, $f_-(t)$ starts only at next-to-leading order 
in the chiral expansion, and that furthermore this contribution is not 
suppressed at $t=0$ for $N_c\to\infty$.

\subsection{Mixing angles}\label{epsilons}

The expressions of the $K_{\ell 3}$ form factors given in the preceding 
subsection involve the $\pi ^0-\eta$ mixing angles
$\stackrel{o}{\varepsilon}, \varepsilon_1, \varepsilon_2$. In defining them
we ignore
isospin breaking through electromagnetic effects, so that the only source
of isospin violation is the quark mass difference $m_d - m_u$.  If
$m_d \neq m_u$, the isosinglet and isovector axial currents
$A_{\mu}^8$ and $A_{\mu}^3$ have nonvanishing off-diagonal matrix
elements between the vacuum and one-meson states.  The $two$ $\pi ^0\-\eta$ 
mixing angles $\varepsilon_1$ and $\varepsilon_2$ are introduced such as to 
define combinations of the axial currents having vanishing off-diagonal matrix elements:
\bea
\langle \Omega |\, \cos \varepsilon_1 \,A_{\mu}^8(0) - \sin
\varepsilon_1 \,A_{\mu}^3(0)\, | \pi^0(p) \rangle &=& 0 \nonumber \\
&& \\
\langle \Omega | \, \cos \varepsilon_2 \, A_{\mu}^3(0) + \sin
\varepsilon_2 \,A_{\mu}^8(0) \,| \eta(p) \rangle &=& 0. \nonumber 
\ena
Both $\varepsilon_1$ and $\varepsilon_2$ are of order ${\cal O}(m_d - m_u)$.
Note that we were not forced to define mixing angles in terms of
matrix elements of the axial currents:  we could have chosen to use
the pseudoscalar densities, or any other operators with the appropriate 
quantum numbers; however, the corresponding expressions
for the mixing angles would in general differ from those given
below. 

Keeping only contributions which are at most linear in the quark mass 
difference $m_d-m_u$, the off-diagonal matrix elements
of the flavor neutral axial currents read
\bea
\langle \Omega |  A_{\mu}^8(0)| \pi^0(p) \rangle &=& i p_{\mu}\, F_{\pi}
\, \varepsilon_1 \nonumber \\
&& \\
\langle \Omega |  A_{\mu}^3(0)| \eta(p) \rangle &=& - i p_{\mu}\, F_{\eta}\,
\varepsilon_2. \nonumber
\ena
The decay constants $F_{\pi}$ and $F_{\eta}$ define the diagonal
matrix elements of the same currents,
\bea
\langle \Omega |  A_{\mu}^3(0)| \pi^0(p) \rangle &=& i p_{\mu} \,
F_{\pi} \nonumber \\ 
&& \\
\langle \Omega |  A_{\mu}^8(0)| \eta(p) \rangle &=& i p_{\mu} \,
F_{\eta}\, \nonumber \;.
\ena
Up to corrections of order ${\cal O}((m_u - m_d)^2)$ that we neglect,  
$F_{\pi}$ can be identified  with the charged pion decay constant, 
$F_{\pi}$ = 92.4 MeV.

At order ${\cal O}({\rm p}^2)$, the two mixing angles coincide and are equal 
to $\stackrel{o}{\varepsilon}$.
At order ${\cal O}({\rm p}^3)$, they both receive ${\cal O}(m_{\rm quark})$
corrections, and are no longer equal.  Explicitly,
\bea
{\varepsilon_1\ +\ \varepsilon_2 \over 2} &=&
\stackrel{o}{\varepsilon} \, + \, 
{1 \over \sqrt{3} R (M_{\eta}^2 - M_{\pi}^2)} \,\bigg\{ {\widehat
m}^3 X_\rho(r)  \nonumber \\
&& -  \bigg({F_K^2 \over
F_{\pi}^2} -1  \bigg) \, \bigg[ {\Delta_{GMO} \over 2} + {1 \over 3}
(r_2 -r) \bigg({r+2 \over r+1 }\bigg) M_{\pi}^2 \bigg] \bigg\}\;,
\label{op3}
\ena
and
\bea
\varepsilon_1 - \varepsilon_2 &=&
\frac{1}{\sqrt{3}R(M_{\eta}^2-M_{\pi}^2)}
\left( \frac{F_K^2}{F_{\pi}^2} - 1 \right)\,
\bigg[ \Delta_{GMO} - \frac{2}{3}(r_2 - r)\left(\frac{r - 2}{r + 1}\right)
M_{\pi}^2 \bigg] \;,
\ena
where we have defined
\be
X_\rho(r) = (r-1)^2 \, [ (3 \rho_1 + 4 \rho_3 )\,(r-1) - \rho_2
\,(r+1)]
\en
in terms of the ${\cal L}_{(0,3)}$ quantities $\rho_{1,2,3}$.

\addtocounter{section}{0}
\setcounter{equation}{0}
\section{\bf Decay rates, $V_{us}$ and $F_K/F_{\pi}$ revisited}
\subsection{\bf The $K^0 \rightarrow \pi^- e^+ \nu$ rate}

For the $K_{e3}$ decay we only need to consider the form factor $f_+(t)$,
and for $K^0 \rightarrow \pi^- e^+ \nu$ Eq.~(\ref{fpK0}) is all we
need.  Ignoring the tiny term in Eq.~(\ref{fpK0}) proportional to
$\stackrel{o}{\varepsilon}$, we may write the decay rate in units of
$10^{-15}~{\mbox{\rm{MeV}}}$ as
\bea
\Gamma[K^0 \rightarrow \pi^- e^+ \nu] &=& V_{us}^2 \bigg[ 105.056
\nonumber \\
&& +\, 203.031\, (r-1)^2\, \bigg(\frac{{\widehat m}^2 A_3}{2} +
\frac{{\widehat m}^2\xi^2}{8}\bigg) \nonumber \\
&& +\, 98.267\, (r-1)^4 \,\bigg(\frac{{\widehat m}^2 A_3}{2} +
\frac{{\widehat m}^2\xi^2}{8}\bigg)^2 \bigg] . \nonumber \\
\label{GammaK0}
\ena
Strictly speaking, the last contribution on the right-hand side of this 
expression, although coming from the square of the ${\cal O}({\rm p}^4)$ 
expression of $f_+^{K^0\pi^-}(t)$, represents an order ${\cal O}({\rm p}^6)$ 
effect in the chiral expansion of the decay rate. However, as can be checked 
{\it e.g.} on Figs. 1 and 2 below, it does not affect our analysis in the 
range of values considered for the Cabibbo angle.

We express ${\widehat m}\xi$ in terms of $V_{us}$ as follows:  first,
we  use Eq.~(\ref{mxiOp4}), but only at lowest order,
\be
{\widehat m}\xi = {1 \over r-1} \bigg({F_K^2 \over F_{\pi}^2} -
1\bigg) \;,
\label{mxi}
\en
and then use the formula for the ratio of the branching rates 
\be
{\Gamma[K \rightarrow \mu \nu] \over \Gamma[\pi \rightarrow \mu \nu]}
= {\left| V_{us}\right|^2 \over \left| V_{ud}\right|^2 } {F_K^2 \over
F_{\pi}^2} {M_{K^+} \over M_{\pi^+}} 
 {\left\{ 1 - (M_{\mu}/M_{K^+})^2 \right\}^2 \over \left\{ 1 -
(M_{\mu}/M_{\pi^+})^2 \right\}^2 } \cdot (1 + \delta_K -
\delta_{\pi})\; ,
\en
where the radiative corrections $\delta_K = 0.0020 \pm 0.0002,
\delta_{\pi} = 0.0017 \pm 0.0002$ nearly cancel \cite{Fink}.

\begin{figure}[ht] 
\centerline{\psfig{figure=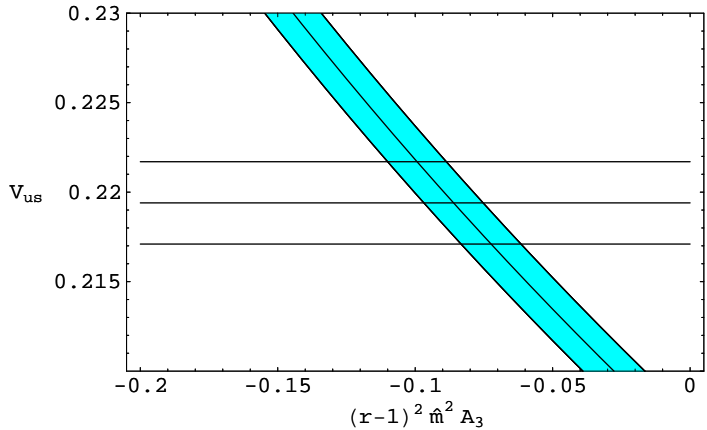,height=5cm}}
\vspace{.5cm}
\caption{The dependence of $V_{us}$ on $A_3$.  Horizontal lines
indicate the range of values for $V_{us}$ given by PDG \protect\cite{PDG98}.}
\label{vusplot}
\end{figure}

Next, we use the experimental values \cite{PDG98} for the branching
rates to fix \cite{lroos} the combination of constants  
\be
{F_K \over F_{\pi}}{V_{us} \over V_{ud}} = 0.2758 \pm 0.0005,
\label{Kmu2pimu2}
\en
together with the unitarity of the Cabibbo-Kobayashi-Maskawa (CKM)
matrix  
\be
V_{ud}^2 + V_{us}^2 = 1.
\label{unitarity}
\en
We ignore $\vert V_{ub}\vert$, which has been obtained \cite{Uraltsev,CLEO99}
from a model-dependent analysis of data from B semileptonic decays to
be (3.25 $\pm$ 0.6) $\times 10^{-3}$; however, even if the central value of 
$V_{ub}$ turned out to be 
three times larger, this would not affect our results.  After
expanding Eq.~(\ref{GammaK0}) in $V_{us}$ around $V_{us} = 0.220$, this
implies 
\bea \label{K0e3rate}
\Gamma[K^0 \rightarrow \pi^- e^+ \nu] &=&  5.390 +
31.10 \,(V_{us} - 0.220)
\nonumber \\
&& +\, (r-1)^2\,{\widehat m}^2 A_3 \,[5.059 + 37.58 \,(V_{us} - 0.220)]
\nonumber \\
&& +\, (r-1)^4\,{\widehat m}^4 A_3^2 \, [1.189 + 10.81\, (V_{us} - 0.220)],
\ena
to be compared with the experimental rate of $4.937 \pm 0.052$.
Consequently, from the knowledge of $V_{us}$ one may extract a value
for the ${\cal L}_{(2,2)}$ quantity ${\widehat m}^2 (r-1)^2 A_3 = (m_s
- {\widehat m})^2 A_3$.  For example, in Figure \ref{vusplot} we
show the band in the ($V_{us}\,,\,A_3$) plane indicated by experiment, 
together
with lines of constant $V_{us}$ corresponding to the Particle Data
Group (PDG) \cite{PDG98} value $V_{us}=0.2196 \pm 0.0023$.  Accepting
this constraint on $V_{us}$ would imply ${\widehat m}^2 (r-1)^2 A_3 =
-0.089 \pm 0.024$.  A naive  dimensional analysis (NDA) \cite{KMSF}
would give 
\be \label{A3NDA}
\vert (r-1)^2\,{\widehat m}^2 \, A_3 \vert\,{\bigg\vert}_{NDA} \approx 
{m_s^2(\Lambda_H)
\over \Lambda_H^2} \approx 0.04 \; ,
\en
where we have taken $m_s(\Lambda_H) \approx$ 200 MeV, and $\Lambda_H \approx$ 1
GeV is a typical hadronic mass scale.

\vspace{1cm}

\subsection{Determination of $V_{ud}$ from nuclear beta decay}

In its 1998 update, PDG \cite{PDG98} recommends for $V_{us}$ only the
value determined from $K_{e3}$ decay, arguing that the value obtained
from hyperon decays (see \cite{DHK} for an early attempt along these
lines) suffers from theoretical uncertainties due to first-order SU(3)
symmetry-breaking effects in the axial-vector couplings. However, within the 
context of the present G$\chi$PT analysis, we may not, without
independent knowledge of $A_3$, use the values for
$V_{us}$ thereby extracted from $K_{e3}$ decay.  Moreover, $F_K$ is
determined from the $K_{\mu 2}$ decay together with the knowledge of $V_{us}$,
and we note that $F_K/F_{\pi}$ appears implicitly in our theoretical
expressions for the $K_{\ell 3}$ form factors.  The origin of this is, of
course, our re-expression Eq.~(\ref{mxi}) of the ${\cal {L}}_{(0,3)}$
constant $\xi$ in terms of $F_K/F_{\pi}$.

>From the unitarity condition for the CKM matrix, it follows that
$|V_{us}|$ may be fixed from the knowledge of the up-down quark-mixing
matrix element of the CKM matrix, $|V_{ud}|$, alone.  The value of
$V_{ud}$ can be determined from several independent sources: nuclear
superallowed Fermi beta decays, free neutron decay, and pion beta
decay.

\begin{figure}[ht] 
\centerline{\psfig{figure=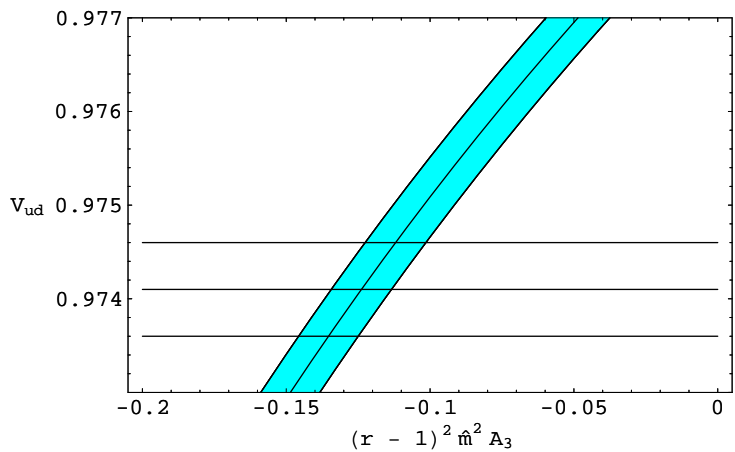,height=5cm}}
\vspace{1cm}
\caption{The dependence of $V_{ud}$ on $A_3$. Horizontal lines
indicate the range of values for $V_{ud}$ determined from superallowed
nuclear beta decay.}
\label{vudplot}
\end{figure}

Currently, superallowed Fermi $0^+ \rightarrow 0^+$ nuclear beta
decays \cite{Ha90,ENAM95,WEIN98}, together with the muon lifetime,
provide the most accurate value,
\be
V_{ud} = 0.9740 \pm 0.0005 .
\label{Vud}
\en 
The precision is limited not by experimental error but by the
estimated uncertainty in theoretical corrections \cite{WEIN98}.  In Figure
\ref{vudplot} we show the band in the $V_{ud}\,,\, A_3$ plane indicated by
experiment, together with lines of constant $V_{ud}$ corresponding to
these values. 

The determinations of $V_{ud}$ from free neutron decay data are approximately 
a factor of four poorer in precision (see the discussion in \cite{WEIN98} and 
references therein),
\be 
V_{ud} = 0.9759 \pm 0.0021 ,
\en
due to the difficulty in separating the vector from the axial piece, but 
planned experiments aiming at an accurate measurement of the electron 
emission asymmetry \cite{deutsch} could change this situation
qualitatively, since the error in this case is primarily of experimental 
origin. 

Finally, the theoretical corrections in the nuclear Fermi transitions are 
absent in the case of the pion beta decay. The present status of this type of 
experiments results in the value \cite{WEIN98}
\be 
V_{ud} = 0.9670 \pm 0.0161.
\en
Here also, the situation might improve in the future \cite{pibeta,deutsch}.

Combining the above results, one obtains
\be 
V_{ud} = 0.9741 \pm 0.0005.
\label{vudnuclear}
\en
Accepting this value for $V_{ud}$ would imply ${\widehat m}^2 (r-1)^2
A_3 = -0.124 \pm 0.022$, somewhat larger than the NDA estimate
Eq.~(\ref{A3NDA}), but still acceptable. 

Finally, we note that recent (model-dependent) analyses of hyperon
semileptonic decays give \cite{Ra96} 
\be 
V_{ud} = 0.9750 \pm 0.0004,
\en
and \cite{Fl98}
\be 
V_{ud} = 0.9743 \pm 0.0009.
\en

\begin{figure}[ht] 
\centerline{\psfig{figure=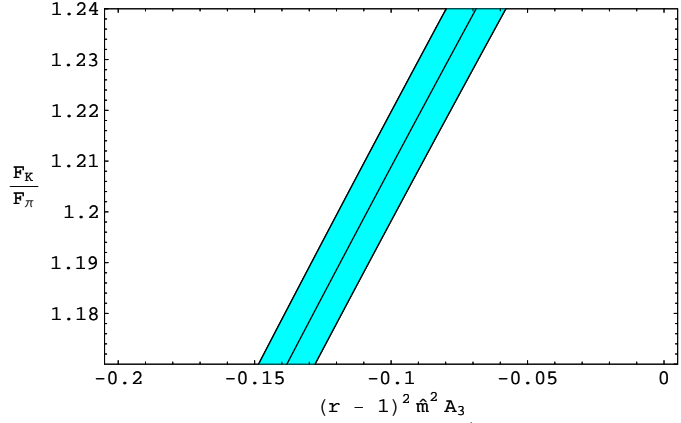,height=4.5cm}}
\vspace{1cm}
\caption{The dependence of $F_K/F_{\pi}$ on $A_3$.}
\label{FKFpiA3}
\end{figure}

Taking the value Eq.~(\ref{vudnuclear}) for $V_{ud}$ ({\it i.e.},
excluding results from hyperon decays), the unitarity relation
Eq.~(\ref{unitarity}) gives 
\be 
V_{us} = 0.2261 \pm 0.0023.
\en
Incorporating the above values for the CKM matrix elements into
Eq.~(\ref{Kmu2pimu2}) then implies 
\be 
\frac{F_K}{F_{\pi}} = 1.189 \pm
0.012 \;\;\; [V_{ud}{\rm ~ from ~nuclear~ beta~ decay]}, 
\label{fkfpinuclear}
\en 
which may be compared with the corresponding result using the PDG
values for $V_{us}$, 
\be 
\frac{F_K}{F_{\pi}} = 1.226 \pm 0.014 \;\;\;
[V_{us}{\rm ~from~ PDG]}.  
\label{fkfpipdg}
\en 
Using Eq.~(\ref{Kmu2pimu2}), which relates $F_K/F_{\pi}$ to $V_{us}$
and $V_{ud}$, Eq.~(\ref{unitarity}) which relates $V_{us}$ and
$V_{ud}$, and Eq.~(\ref{K0e3rate}) which relates $V_{us}$ and
${\widehat m}^2 (r-1)^2 A_3$, we may directly relate $F_K/F_{\pi}$ and
${\widehat m}^2 (r-1)^2 A_3$, see Figure \ref{FKFpiA3}.

\subsection{$m_u \neq m_d$ effects on the $K^+ \rightarrow \pi^0 e^+ \nu$
rate} 

We may treat the decay of the $K^+$ similarly; the result is most
conveniently expressed in terms of the ratio

\bea \label{rates}
{\Gamma[K^+ \rightarrow \pi^0 e^+ \nu] \over \Gamma[K^0 \rightarrow
\pi^- e^+ \nu] }&=&( R^{+0})^2 \, \bigg[ 0.49615 - 0.00240 \, (r-1)^2\,
\bigg(\frac{{\widehat m}^2 A_3}{2} + \frac{{\widehat
m}^2\xi^2}{8}\bigg) \nonumber \\ 
&& +\, 0.00234\, (r-1)^4 \,\bigg(\frac{{\widehat m}^2 A_3}{2} +
\frac{{\widehat m}^2\xi^2}{8}\bigg)^2 \bigg] , \nonumber \\
\ena
where $R^{+0}$ is the ratio
\be
R^{+0}  \equiv {{\displaystyle
f_{+}^{K^+\pi^0}(0)}\over{\displaystyle f_{+}^{K^0\pi^-}(0)}}
\label{Rp0}\;.
\en
Note that the last two terms in the brackets in Eq.~(\ref{rates}) arise
from the small differences in phase space which are due to mass
differences between $K^0$ and $K^+$, and between $\pi ^+$ and $\pi
^0$.   

Proceeding as we did for the $K^0$ rate above, {\it i.e.}, using
Eqs. (\ref{mxi}) and (\ref{Kmu2pimu2}), we obtain
\bea
{\Gamma[K^+ \rightarrow \pi^0 e^+ \nu] \over \Gamma[K^0 \rightarrow
\pi^- e^+ \nu] }&=&( R^{+0})^2 \, \big\{ 0.4961 + 0.0040 \,(V_{us} -
0.22) 
\nonumber \\
&& - (r-1)^2\,{\widehat m}^2 \, A_3 \,[0.0011 + 0.0041 \,(V_{us} -
0.22)] \nonumber \\ 
&& +   0.0005\, (r-1)^4 \,{\widehat m}^4 \,A_3^2 \big\}\;.
\ena
The correction terms in the curly brackets are completely negligible
for $A_3$ and $V_{us}$ in the range determined above, since they are
at most of the order of 0.0001, compared to the leading term 0.4961.

The measured rates may be deduced from the data given by the Particle
Data Group \cite{PDG98}:
\bea
\Gamma[K^+ \rightarrow \pi^0 e^+ \nu]_{exp} &=& 2.561 \pm 0.032 \nonumber \\
\Gamma[K^0 \rightarrow \pi^- e^+ \nu]_{exp} &=& 4.937 \pm 0.052, \nonumber \\
\ena
where, as above, we express all rates in units of $10^{-15}$ MeV; consequently,
\be
{\Gamma[K^+ \rightarrow \pi^0 e^+ \nu]_{exp} \over \Gamma[K^0 \rightarrow
\pi^- e^+ \nu]_{exp} } =  0.519 \pm 0.016.
\en
Consequently, the experimental value for $R^{+0}$ is 1.023 $\pm$ 0.016.


Up to and including terms of order ${\cal O}({\rm p}^4)$, the theoretical
expression for $R^{+0}$ is given by Eq.~(\ref{fKpfK0}); the terms
proportional to ${\widehat m}^2\xi^2 (r-1)^2$ and to ${\widehat m}^2C_2^P$ 
are of order ${\cal O}({\rm p}^4)$. We first estimate the 
${\cal O}({\rm p}^3)$ correction terms
in Eq.~(\ref{op3}) by NDA.  For the first one, we obtain 
\be 
\left|{{\widehat m}^3 X_\rho(r) \over
M_{\pi}^2}\right|_{NDA} 
\approx \left|{8\,r^3 {\widehat m}^3 \rho \over
M_{\pi}^2}\right|_{NDA} \approx {8\,m_s^3(\Lambda_H) \over M_{\pi}^2 \Lambda_H}
\approx  3.2, 
\en
where $\rho$ characterizes the size of the ${\cal L}_{(0,3)}$ low-energy 
constants $\rho_{\alpha},\,\alpha =
1,2,3$.  This implies (using $R=43.5$, from reference \cite{GLPR})
\be \label{m3Xrho}
{{\widehat m}^3 X_\rho(r) \over R (M_{\eta}^2 - M_{\pi}^2)}
\ \approx \ {0.22 \over R} \ \approx \ 0.005. 
\en
Similarly, the last term in Eq.~(\ref{op3}) is
\bea
{1 \over \sqrt{3} R (M_{\eta}^2 - M_{\pi}^2)} \bigg( {F_K^2 \over
F_{\pi}^2} -1  \bigg) \, \bigg[ {\Delta_{GMO} \over 2} + {1 \over 3}
(r_2 -r) \bigg({r+2 \over r+1 }\bigg) M_{\pi}^2 \bigg] \quad &&
\nonumber \\
\approx \quad {-0.033 \over R} + {0.006 (r_2 - r) \over R} \,
\bigg({r+2 \over r+1 } \bigg) && \\
\qquad \approx \;
-0.00075 + 0.00014 \,(r_2 -r)\, \bigg({r+2 \over r+1 } \bigg)\;,
\nonumber && 
\ena
which is quite negligible. 

Finally, we turn to the ${\cal O}({\rm p}^4)$ corrections to $R^{+0}$. From 
\be
{\widehat m}^2\xi^2 (r-1)^2  \approx \bigg( {F_K^2 \over F_{\pi}^2}
-1  \bigg)^2  \approx 0.24,
\en
we obtain 
\be \label{mxi2eps0}
{{\widehat m}^2\xi^2 (r-1)^2  \stackrel{o}{\varepsilon} \over \sqrt{3}}
\approx 0.14 \stackrel{o}{\varepsilon} \;.
\en
The NDA estimate for the $C_2^P$ term gives
\be
{\widehat m}^2 \,\left| C_2^P \right| \,(r-1)^2\,\bigg[ {1\over R}\ +\
{{2\stackrel{o}{\varepsilon}}\over{\sqrt{3}}}\bigg] \approx 0.04
\,\cdot \bigg[ {1\over R}\ +\
{{2\stackrel{o}{\varepsilon}}\over{\sqrt{3}}}\bigg] \;,
\en
where we note that no assumption has been made here of any suppression
of $C_2^P$ due to Zweig rule violation.  We have verified that ${\cal O}({\rm
p}^4)$ corrections to $\varepsilon_1\ +\ \varepsilon_2$ are negligibly
small.
Using the numerical form of Eq.~(\ref{eps0}),
\be
\stackrel{o}{\varepsilon}\,=\, {1 \over R} \,\bigg[0.533  + 0.502 \, {r - 1
\over r+1}\, \bigg(1 - {r \over r_2} \bigg) \bigg],
\en
the upshot is that $R^{+0}$ may be written (recall Eq.~(\ref{fKpfK0}))
\bea \label{R-vs-r}
R^{+0} &=& 1 + \sqrt{3}\,\stackrel{o}{\varepsilon} - \,
0.14\,\stackrel{o}{\varepsilon} \pm \, {0.22 \over R}  \;, 
\ena
where the term $- 0.14\,\stackrel{o}{\varepsilon}$ comes from
Eq.~(\ref{mxi2eps0}); the indicated uncertainty, coming from other
higher-order corrections, is dominated by the estimation of the
contributions from the ${\cal L}_{(0,3)}$ terms (see Eq.~(\ref{m3Xrho})
above). Numerically, this implies 
\be
R^{+0} \,=\,   1  + {1 \over R}\, \bigg[0.85 + 0.80  \,
{r - 1 \over r+1}\, \bigg(1 - {r \over r_2} \bigg) \bigg] \pm {0.22
\over R},
\en
and taking, for example, the commonly accepted value \cite{GLPR} $R$ =
43.5 $\pm$ 2.2,  
\be
R^{+0} = 1.020 \pm 0.005 \ + \ 0.018  \, {r - 1
\over r+1}\, \bigg(1 - {r \over r_2} \bigg),
\en
which is consistent, for any permissible value of $r$, with the
experimental value 1.023 $\pm$ 0.016 which we deduced above.  In view
of this experimental uncertainty, we will leave for a later time the
careful investigation of the ${\cal O}({\rm p}^4)$ effects mentioned above,
together with the analysis of electromagnetic corrections (the radiative 
corrections to $R^{+0}$ in the standard case have been investigated in 
Ref. \cite{neurup}).  Note that
some (but not all) of the existing experimental data has been
published with radiative corrections, but often without mention of how
these corrections have been implemented \cite{Nefkens}.  More precise
knowledge of $R^{+0}$ would be useful in constraining the relationship
displayed in Eq.~(\ref{R-vs-r}) between the two quark-mass ratios $R$
and $r$, thereby testing the relevance of G$\chi$PT.

\addtocounter{section}{0}
\setcounter{equation}{0}
\section{Form factor slopes}\label{f0} 

\begin{figure}[ht] \label{lambda0plot}
\centerline{\psfig{figure=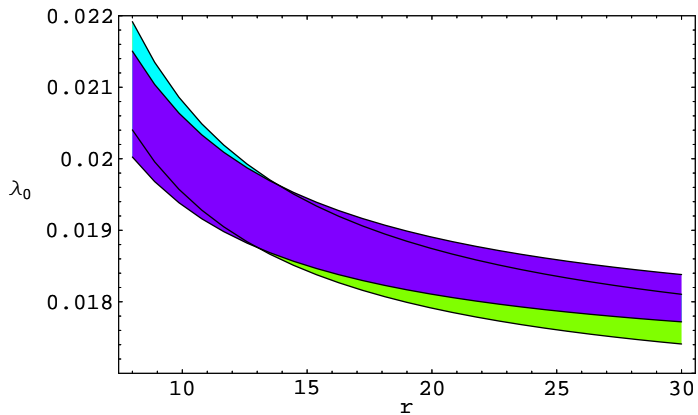,height=5cm}}
\vspace{1.5cm}
\caption{The dependence of $\lambda_0$ on $r$.  The darker shaded
region indicates the range of values Eq.~~(\protect\ref{fkfpinuclear}) for
$\lambda_0$ determined by using $V_{ud}$ from superallowed nuclear
beta decay, while the lighter shaded region shows the corresponding
range Eq.~(\protect\ref{fkfpipdg})  determined by using the PDG values for
$V_{us}$.  See text for details.} 
\end{figure}

Analyses of $K_{\ell 3}$ data frequently assume a linear dependence
\be
f^{K\pi}_{+,0} (t) = f^{K\pi}_+ (0) \left[ 1 + \lambda_{+,0}
\frac{t}{M^2_{\pi^+}} \right]
\label{s38}\;,
\en
where, as usual, the scalar form factor is defined as
\be
f^{K\pi}_0(t) = f^{K\pi}_+(t) + {t \over M_K^2 - M_{\pi}^2}
f^{K\pi}_-(t)\;. 
\en
The parameter $\lambda_+^{K\pi}$ is identical to that of S$\chi$PT,
\bea
{\lambda_+^{K\pi}\over M_{\pi^+}^2}  &=& {1\over 6}\,\langle\,{\rm
r}^2\,\rangle^{\pi}_{\rm V}\nonumber \\
&& + 
\ {1\over{64\pi^2F_0^2}}\,\bigg[ 1\,-\,{1\over 2}h_1\big(
{{M_{\pi}^2}\over{M_{K}^2}}\big)\,-\,{1\over 2}h_1\big(
{{M_{K}^2}\over{M_{\eta}^2}}\big)\,+\,{5\over 12}\ln \big(
{{M_{\pi}^2}\over{M_{K}^2}}\big)\,+\,{1\over 4} \ln \big(
{{M_{K}^2}\over{M_{\eta}^2}}\big) \bigg]\ ,
\ena
with
\be
h_1(x)\ +\  {1\over 2}\,{{(x^3-3x^2-3x+1)}\over{(x-1)^3}}\,\ln x\,+
\,{1\over 2}\,\left({{x+1}\over{x-1}}\right)^2\,-\,{1\over 3}\ ,
\en
and
\be
F_{\pi}^2 \, \langle\,{\rm r}^2\,\rangle^{\pi}_{\rm V} \,\ =\,
12\,L_9^r - \frac{1}{32\pi^2}\bigg\{2\ln\frac{M_\pi^2}{\mu^2} + 
\ln\frac{M_K^2}{\mu^2} + 3\bigg\} \ .
\en
On the other hand, the G$\chi$PT expression for $\lambda_0^{K\pi}$,
\bea \label{lambda0}
{\lambda_0^{K\pi} \over M_{\pi^+}^2} &=& {1\over 2}\bigg[
\left({{F_K^2}\over{F_{\pi}^2}}\,-\,1 \right)\,-\,{1\over
  2}\left({{F_K^2}\over{F_{\pi}^2}}\,-\,1\right)^2\,- \,{\widehat
  m}^2\,(r-1)^2\,A_3\,{r+3 \over r-1}\bigg] {1 \over M_K^2 - M_{\pi}^2 } \nonumber\\ 
&-& {1\over{384\pi^2F_0^2}}\bigg[ 5\,-
\,{{4 M_{\pi}^2 \,(r + 1)\,
\hat{\epsilon}(r)}\over{M_K^2+M_{\pi}^2}}\bigg]\,h_2\big( 
{{M_{\pi}^2}\over{M_{K}^2}}\big)\\ 
&-& {1\over{384\pi^2F_0^2}}\bigg[ 3\,+ \nonumber \\
&&
\,2\,\left({{M_{\eta}^2-M_K^2}\over{M_{\eta}^2+M_K^2}}\right)
\left({{M_K^2+M_{\pi}^2}\over{M_K^2-M_{\pi}^2}}\right)
\left(1-{{\Delta_{GMO}\,(r+3)}\over{2 \, (M_K^2+M_{\pi}^2)\, (r-1)}}
\right)\bigg]\,h_2\big(
{{M_{K}^2}\over{M_{\eta}^2}}\big)\nonumber\\
&+& {1\over{64\pi^2F_0^2}}\ ,\nonumber
\ena
where
\be
h_2(x)\ =\ {3\over 2}\,
\left({{1+x}\over{1-x}}\right)^2\,+\,{{3x(1+x)}\over{(1-x)^3}}\,\ln x\ ,
\en
differs from the S$\chi$PT result.
This expression for $\lambda_0$ explicitly displays the dependence on
${\widehat m}^2 (r-1)^2 A_3$ and on $r$, while its dependence on
$V_{us}$ is implicit in the $F_K/F_{\pi}$ terms.  However, as we saw
above in Figure \ref{FKFpiA3}, $F_K/F_{\pi}$ and ${\widehat m}^2
(r-1)^2 A_3$ are correlated once we know the rate for $K^0 \rightarrow
\pi^- e^+ \nu$.  Therefore, for any choice of the quark mass ratio
$r$, Eq.~(\ref{lambda0}) gives a range of values for $\lambda_0$
corresponding to a given region in the ($F_K/F_{\pi} ,\ {\widehat m}^2
(r-1)^2 A_3$) plane.  For example, Eq.~(\ref{fkfpinuclear}) defines one
region in Figure \ref{FKFpiA3}, while Eq.~(\ref{fkfpipdg}) defines
another.  We display in Figure \ref{lambda0plot} the corresponding
regions in the ($\lambda_0, r$) plane.

This analysis was done assuming that the Zweig-violating parameter
$Z_0^S$=0.  The whole dependence on $Z_0^S$ is contained (through
$\zeta = Z_0^S/A_0$) in the parameter $\hat{\epsilon}(r)$.  As pointed
out in \cite{KS}, the vacuum stability requirement $B_0 \ge 0$
implies a upper bound on $\zeta$, which yields
\be
\hat{\epsilon}(r) - \hat{\epsilon}(r)|_{\zeta = 0} \ < \ (1 -
\hat{\epsilon}(r)|_{\zeta = 0})\ {2 \over r+2}
\en
We find that taking $\zeta = 0$ or its maximal allowed value makes a
difference of less than 0.0004 in $\lambda_0$, for any choice of $r$.

In S$\chi$PT, including one-loop corrections, the result \cite{GL85b} is
\be \lambda_0 = 0.017 \pm 0.004, \en where the error is an estimate of
the uncertainties due to higher-order contributions.  The experimental
situation remains unclear, in view of the inconsistency between 
some more recent data \cite{PDG98} and the result 
($\lambda_0 = 0.019 \pm 0.004$) of
the high-statistics experiment of Donaldson {\it et al}
\cite{donaldson}.


\addtocounter{section}{0}
\setcounter{equation}{0}
\section{Summary and discussion}

In the preceding pages, we have studied the $K_{\ell 3}$ form factors at 
${\cal O}({\rm p}^4)$ precision within the generalized framework of chiral 
perturbation theory. In this connection, three (related) issues have been 
discussed: the extraction of the CKM matrix element $V_{us}$ from the 
experimental value of the $K_{\ell 3}$ branching ratios, the determination of 
the ratio $F_K/F_\pi$ of pseudoscalar decay constants, and the prediction for 
the slope $\lambda_0$ of the scalar form factor.

In the Leutwyler-Roos analysis \cite{lroos}, the expression of 
$f_+^{K^0\pi^-}(0)$ is written as
\be\lbl{LRexp}
f_+^{K^0\pi^-}(0) = 1 + f_1 + f_2 +\cdots\,,
\en
where the
contributions $f_1$ and $f_2$ are of order $m_{\rm quark}$ and $m_{\rm
quark}^2$ respectively. The one-loop  S$\chi$PT contribution $f_1$ arises 
from Goldstone boson loops only, a counterterm contribution at this order of 
the standard counting being forbidden by the Ademollo-Gatto theorem. This 
leads to a parameter free prediction $f_1 = -0.023$. 
Furthermore, $f_2$, which would arise at chiral order ${\cal O}({\rm p}^6)$ 
in S$\chi$PT, has been estimated in~\cite{lroos} to be $-$0.016 $\pm$ 0.008 
by using a model for pion and kaon wave functions to compute matrix elements 
in the infinite momentum frame. This leads to a value 
$V_{us}= 0.2196\pm0.0023$. Assuming unitarity of the CKM matrix in a three 
generation standard model, and using the existing estimates of 
$\vert V_{ub}\vert$, the determination $V_{ud} = 0.9741\pm 0.0005$ from 
superallowed nuclear $\beta$-decays leads instead to $V_{us}= 0.2261\pm0.0022$.
Therefore, unitarity of the CKM matrix can only be restored at the expense of 
having the S$\chi$PT two-loop correction $f_2$ at least as large as the one-loop 
contribution $f_1$.
This is far from signaling a failure of the chiral expansion in the present 
case, since $f_1$ might be anomalously small, being, for instance, 
suppressed, even compared to $f_2$, in the large-$N_c$ limit (another 
consequence of the Ademollo-Gatto theorem). In G$\chi$PT, the corresponding 
expansion of $f_+^{K^0\pi^-}(0)$ reads
\be\lbl{LRseries}
f_+^{K^0\pi^-}(0) = 1 + {\tilde f}_1 + {\tilde f}_2 +\cdots\,,
\en
where ${\tilde f}_1$ collects all ${\cal O}({\rm p}^4)$ contributions in the 
generalized chiral counting, and contains ${\cal O}(m_{\rm quark}^2)$ terms. 
The difference between ${\tilde f}_1$ and $f_1$
\be
{\tilde f}_1 - f_1 = \bigg[ {1\over 8}{\widehat m}^2\,\xi^2\,+
\,{1\over 2}{\widehat m}^2A_3\bigg]\,(r-1)^2\,\big(1+{1\over R}\big) \;,
\en
involves a single low-energy constant from ${\cal L}_{(2,2)}$, $A_3$, which 
would appear only at order ${\cal O}({\rm p}^6)$ in S$\chi$PT. 
The contribution of the ${\cal L}_{(2,1)}$ low-energy constant $\xi$ is 
determined by the value of the ratio $F_K/F_\pi$. The value of 
$V_{ud}$ from the nuclear $\beta$-decay data can be accommodated by 
\be
(r-1)^2{\widehat m}^2A_3 = -0.124\pm 0.022\,,
\quad\frac{F_K}{F_\pi} = 1.189\pm 0.012\,.
\en
The corresponding difference is ${\tilde f}_1 - f_1 = -0.04\pm 0.01$, and 
it represents  the value which would be required in the Leutwyler-Roos 
analysis for the two-loop contribution $f_2$ (instead of the estimate 
$f_2=-0.016\pm 0.008$ of \cite{lroos}) in 
order to maintain the unitarity of the CKM matrix. Of course, a confirmation, 
with a comparable accuracy,  
from other sources (neutron decay, pion $\beta$-decay) of the value of 
$V_{ud}$ obtained from nuclear $\beta$-decays can only be welcome. 

This determination of $A_3$ and the decrease in the value of the ratio of 
decay constants, as compared to the number $F_K/F_\pi=1.22\pm 0.01$
\cite{lroos}, is compatible with the present experimental information 
concerning the difference in the $K^0_{e3}$ and $K^+_{e3}$ decay rates, and 
induces only a mild modification in the prediction for the slope of the 
scalar form factor $\lambda_0$, which, as a function of the quark mass ratio 
$m_s/{\widehat m}$, varies between 0.0018 and 0.0022,  well 
within the range set by the high-statistics $K^0_L$ experiment of Donaldson 
{\it et al.} \cite{donaldson}. The higher values of $\lambda_0$ obtained by 
some of the more recent experiments \cite{PDG98} are therefore difficult to 
understand at the theoretical level, and cannot be ascribed, within 
G$\chi$PT, to the manifestation of a smaller value of the bilinear light 
quark condensate.

We thus conclude that the nuclear $\beta$-decay determination of
$V_{ud} = 0.9741\pm 0.0005$ need not be in contradiction with the present
values of the $K_{e3}$ decay rates and with chiral perturbation theory. 
One should then ask how
the corresponding increase of $\vert V_{us}\vert$ by about 2.5 standard 
deviations would
manifest itself in various observables. We have already mentioned that the
present understanding of hyperon semi-leptonic decays is compatible with the
suggested update of $V_{us}$. The effect on the $K_{e4}$ decay rates should
be analyzed separately \cite{FKSunpub}. Some effect on the
extraction of the ${\cal O}({\rm p}^4)$ low-energy constants $L_1 , L_2 $
and $L_3$ is to be expected {\it a priori}, but a precise statement requires 
a closer analysis. Finally, it is worth mentioning the possible effect on 
hadronic spectral functions which are extracted from the decays
$\tau \rightarrow{\rm hadrons} + \nu_{\tau}$ and used for a determination of
fundamental QCD parameters \cite{aleph1,pich,aleph2}. While the
non-strange ($\bar ud$) spectral functions should be only barely affected by
an increase of $\sim 0.25\%$, the recently published \cite{aleph2} strange
($\bar us$) spectral functions should be reduced by $\sim 4.6\%$.
Consequently, we would expect no notable influence on the determination of
$\alpha_S(M_{\tau}^2)$ \cite{aleph1}, whereas the central value of the 
running strange quark mass ${\rm m_s}$ determined recently 
\cite{pich,aleph2} could increase by $\sim 15-20\%$. 
The issue certainly deserves a more detailed study.

\indent

\noindent
{\bf Acknowledgments}

Useful correspondence with W.S. Woolcock, W.C. Haxton and J. Comfort is 
kindly acknowledged. 
We also thank N. Vinh Mau for clarifying discussions.

\vfill

\newpage

\begin{center}{\Large \bf Appendices}\end{center}

\newcounter{zahler}
\addtocounter{zahler}{1}
\renewcommand{\thesection}{\Alph{zahler}}
\renewcommand{\theequation}{\Alph{zahler}.\arabic{equation}}

\section{\bf Expansion and renormalization of the effective
Lagrangian}\label{knecht} 

In this Appendix, we display the structure of the effective action of 
G$\chi$PT up to order ${\cal O}({\rm p}^4)$. For a general discussion of the 
G$\chi$PT expansion framework, we refer the reader to the existing 
literature \cite{FSS,SSF,KS}. 

At leading order, the generalized expansion is described by
${\tilde{\cal L}}^{(2)}$, which was first given in  Ref.~\cite{FSS}:
\bea
{\tilde{\cal L}}^{(2)}&=&{1\over 4}F_0^2
\left\{\langle D_\mu U^+D^\mu U\rangle +2B_0\langle
U^+\chi+\chi^+U\rangle \right.\nonumber\\
&&\qquad + A_0\langle (U^+\chi)^2+(\chi^+U)^2\rangle
+Z_0^S\langle U^+\chi+\chi^+U\rangle ^2\lbl{glead} \lbl{L2}\\
&&\qquad + Z_0^P\langle U^+\chi-\chi^+U\rangle ^2
+\left. H_0\langle \chi^+\chi\rangle\right\}\ .\nonumber
\ena
The notation is as in Refs.  \cite{GL85a,GL85b}, except for the consistent
removal of the factor $2B_0$ from $\chi$, the parameter that collects 
the scalar and pseudoscalar sources,
\eq
\chi={s}+i{ p}={\cal M}+\cdots \ ,\, {\cal M}={\rm diag}(m_u,m_d,m_s)
\ .\lbl{chi}
\en
In G$\chi$PT, the next-to-leading-order corrections are of order 
${\cal O}({\rm p}^3)$, and still 
occur at tree level only. They are
embodied in ${\tilde{\cal L}}^{(3)}={\cal L}_{(2,1)}+{\cal L}_{(0,3)}$, 
which reads  \cite{SSF,KMS}
\bea
\tilde{\cal L}^{(3)}&=&{1\over 4}F_0^2
\left\{\xi\langle D_\mu U^+D^\mu U(\chi^+U+U^+\chi)\rangle 
+\tilde\xi\langle D_\mu U^+D^\mu U\rangle
\langle\chi^+U+U^+\chi\rangle\right.\nonumber\\
&&\qquad + \rho_1\langle (\chi^+U)^3+(U^+\chi)^3\rangle
+\rho_2\langle (\chi^+U+U^+\chi)\chi^+\chi\rangle\nonumber\\
&&\qquad + \rho_3\langle\chi^+U-U^+\chi\rangle
\langle(\chi^+U)^2-(U^+\chi)^2\rangle\lbl{L3} \\
&&\qquad + \rho_4\langle(\chi^+U)^2
+(U^+\chi)^2\rangle
\langle\chi^+U+U^+\chi\rangle\nonumber\\
&&\qquad + \rho_5\langle\chi^+\chi\rangle
\langle\chi^+U+U^+\chi\rangle\nonumber\\
&&\qquad + \left.\rho_6\langle\chi^+U-U^+\chi\rangle^2
\langle\chi^+U+U^+\chi\rangle+\rho_7\langle\chi^+U+U^+\chi\rangle^3
\right\}\ .\nonumber
\ena
The tree-level contributions at order ${\cal O}({\rm p}^4)$ are contained 
in \footnote{Contributions from the odd intrinsic parity sector are also 
present 
in the effective Lagrangian; at order ${\cal O}({\rm p}^4)$ they are given by 
the Wess-Zumino term and are the same for S$\chi$PT and for G$\chi$PT, so that
we do not display them here.}
\eq
\tilde{\cal L}^{(4)}={\cal L}_{(4,0)}+{\cal L}_{(2,2)}+{\cal L}_{(0,4)}+
B_0^2{\cal L}'_{(0,2)} + B_0{\cal L}'_{(2,1)} + B_0{\cal L}'_{(0,3)}\ .
\lbl{L4}
\en
The part without explicit chiral symmetry breaking, ${\cal L}_{(4,0)}$, 
is described by the same low-energy constants $L_1$, $L_2$, $L_3$,
$L_9$ and $L_{10}$ as in S$\chi$PT  \cite{GL85a}:
\bea
{\cal L}_{(4,0)}&=& 
L_1 \langle D_{\mu}U^+D^{\mu}U\rangle \langle
D_{\nu}U^+D^{\nu}U\rangle + L_2 \langle D_{\mu}U^+D^{\nu}U\rangle
\langle D_{\mu}U^+D^{\nu}U\rangle \nonumber\\
& & + L_3 \langle D_{\mu}U^+D^{\mu}U D_{\nu}U^+D^{\nu}U\rangle -
i L_9 \langle F_{\mu\nu}^RD^{\mu}UD^{\nu}U^+ 
               + F_{\mu\nu}^LD^{\mu}U^+D^{\nu}U\rangle\\
& & + L_{10} \langle U^+F_{\mu\nu}^RUF^{L\mu\nu}\rangle +
H_1 \langle F_{\mu\nu}^RF^{R\mu\nu} + F_{\mu\nu}^LF^{L\mu\nu} \rangle
\nonumber \;.
\ena
The new term ${\cal L}_{(2,2)}$ would count as ${\cal O}({\rm p}^6)$ in
S$\chi$PT, and is given by:  
\bea \label{L22}
{\cal L}_{(2,2)}&=& {1\over{4}}F_0^2\left\{
A_1 \langle D_{\mu}U^+D^{\mu}U (\chi^+\chi + U^+\chi\chi^+U)\rangle
\right. \nonumber\\    
& &\qquad + A_2 \langle D_{\mu}U^+U\chi^+D^{\mu}UU^+\chi\rangle
\nonumber\\ 
& &\qquad + A_3 \langle
D_{\mu}U^+U(\chi^+D^{\mu}\chi-D^{\mu}\chi^+\chi) +  D_{\mu}UU^+(\chi
D^{\mu}\chi^+ - D^{\mu}\chi\chi^+)\rangle \nonumber\\ 
& &\qquad + A_4 \langle
D_{\mu}U^+D^{\mu}U\rangle\langle\chi^+\chi\rangle  \nonumber\\
& &\qquad + B_1 \langle D_{\mu}U^+D^{\mu}U (\chi^+U\chi^+U + U^+\chi
                        U^+\chi)\rangle \nonumber\\ 
& &\qquad + B_2 \langle D_{\mu}U^+\chi D^{\mu}U^+\chi + 
                        \chi^+D_{\mu}U\chi^+D^{\mu}U\rangle\nonumber\\ 
& &\qquad + B_3 \langle U^+D_{\mu}\chi U^+D^{\mu}\chi +
                       D_{\mu}\chi^+UD^{\mu}\chi^+U\rangle \nonumber\\
& &\qquad + B_4 \langle
D_{\mu}U^+D^{\mu}U\rangle\langle\chi^+U\chi^+U+U^+\chi U^+\chi\rangle\nonumber\\
& &\qquad + C_1^S \langle D_{\mu}U\chi^+ + \chi D^{\mu}U^+\rangle
            \langle D_{\mu}U\chi^+ + \chi D^{\mu}U^+\rangle \\
& &\qquad + C_2^S \langle D_{\mu}\chi^+U + U^+D^{\mu}\chi\rangle
            \langle D_{\mu}U^+\chi + \chi^+D^{\mu}U\rangle \nonumber\\
& &\qquad + C_3^S \langle D_{\mu}\chi^+U + U^+D^{\mu}\chi\rangle
            \langle D_{\mu}\chi^+U + U^+D^{\mu}\chi\rangle \nonumber\\
& &\qquad + C_1^P \langle D_{\mu}U\chi^+ - \chi D^{\mu}U^+\rangle
            \langle D_{\mu}U\chi^+ - \chi D^{\mu}U^+\rangle \nonumber\\
& &\qquad + C_2^P \langle D_{\mu}\chi^+U - U^+D^{\mu}\chi\rangle
            \langle D_{\mu}U^+\chi - \chi^+D^{\mu}U\rangle \nonumber\\
& &\qquad + C_3^P \langle D_{\mu}\chi^+U - U^+D^{\mu}\chi\rangle
            \langle D_{\mu}\chi^+U - U^+D^{\mu}\chi\rangle \nonumber\\
& &\qquad + D^S \langle D_{\mu}U^+D^{\mu}U(\chi^+U+U^+\chi)\rangle 
                               \langle\chi^+U+U^+\chi\rangle \nonumber\\
& &\qquad + D^P \langle D_{\mu}U^+D^{\mu}U(\chi^+U-U^+\chi)\rangle 
                               \langle\chi^+U-U^+\chi\rangle \nonumber\\
& &\qquad + \left. H_2 \langle D_{\mu}\chi^+D^{\mu}\chi\rangle
\right\}\nonumber \;.
\ena
Notice that the number of counterterms (17) involved in ${\cal L}_{(2,2)}$ 
agrees with Refs. \cite{FeaSch,BCE1999}. However, in both cases, different 
bases have been used.
Finally, the tree-level contributions which behave as 
${\cal O}(m_{\rm quark}^4)$ in 
the chiral limit are contained in ${\cal L}_{(0,4)}$,
\bea
{\cal L}_{(0,4)} &=& {1\over{4}}F^2_0\left\{
E_1 \langle (\chi^+U)^4 + (U^+\chi)^4 \rangle \right.\nonumber\\
& &\qquad + E_2 \langle \chi^+\chi(\chi^+U\chi^+U+U^+\chi U^+\chi) \rangle
                                                     \nonumber\\
& &\qquad + E_3 \langle \chi^+\chi U^+\chi\chi^+U \rangle \nonumber\\
& &\qquad +
F_1^S \langle \chi^+U\chi^+U + U^+\chi U^+\chi \rangle^2 \nonumber\\
& &\qquad +
F_2^S \langle (\chi^+U)^3 + (U^+\chi)^3 \rangle\langle \chi^+U + U^+\chi\rangle
\nonumber\\
& &\qquad + 
F_3^S \langle \chi^+\chi(\chi^+U + U^+\chi)\rangle
                           \langle\chi^+U + U^+\chi\rangle\nonumber\\
& &\qquad + 
F_4^S \langle (\chi^+U)^2 + (U^+\chi)^2 \rangle\langle \chi^+\chi \rangle
\nonumber\\
& &\qquad +
F_1^P \langle \chi^+U\chi^+U - U^+\chi U^+\chi \rangle^2 \\
& &\qquad + 
F_2^P \langle (\chi^+U)^3 - (U^+\chi)^3 \rangle\langle \chi^+U - U^+\chi\rangle
\nonumber\\
& &\qquad + 
F_3^P \langle \chi^+\chi(\chi^+U - U^+\chi)\rangle
                           \langle\chi^+U - U^+\chi\rangle \nonumber\\
& &\qquad +
F_5^{SS} \langle (\chi^+U)^2 + (U^+\chi)^2 \rangle
         \langle \chi^+U + U^+\chi \rangle ^2 \nonumber\\
& &\qquad + 
F_6^{SS} \langle \chi^+\chi \rangle \langle \chi^+U + U^+\chi \rangle ^2 
\nonumber\\
& &\qquad +
F_5^{SP} \langle (\chi^+U)^2 + (U^+\chi)^2 \rangle
         \langle \chi^+U - U^+\chi \rangle ^2 \nonumber\\
& &\qquad +
F_6^{SP} \langle \chi^+\chi \rangle \langle \chi^+U - U^+\chi \rangle ^2 
\nonumber\\
& &\qquad + 
F_7^{SP} \langle (\chi^+U)^2 - (U^+\chi)^2 \rangle
         \langle \chi^+U - U^+\chi \rangle \langle \chi^+U + U^+\chi \rangle
\nonumber\\
& &\qquad + \left. 
H_3 \langle \chi^+\chi\chi^+\chi \rangle +
H_4 \langle \chi^+\chi \rangle ^2 \right \} \nonumber \;.
\ena
Contributions from ${\cal L}_{(0,4)}$ would only appear at order 
${\cal O}({\rm p}^8)$ in S$\chi$PT, which, to the best of our knowledge, 
have not been discussed in the literature.

The ${\cal O}({\rm p}^4)$ loop corrections to the processes studied here 
involve only graphs with one or
two vertices from ${\tilde{\cal L}}^{(2)}$:
\eq
{\tilde{\cal Z}}^{(4)}_{\rm 1\, loop} = {\tilde{\cal Z}}^{(4)}_{\rm tadpole}+
{\tilde{\cal Z}}^{(4)}_{\rm unitarity}+\cdots \lbl{Zloop}
\en
The divergent parts of these one-loop graphs have been subtracted at a scale
$\mu$ in the same dimensional renormalization scheme as described in
 \cite{GL85a}. Accordingly, the low energy constants of ${\cal L}_{(4,0)}$,
${\cal L}_{(2,2)}$, and ${\cal L}_{(0,4)}$ stand for the renormalized 
quantities, with an explicit logarithmic scale dependence ($X(\mu )$ denotes
generically any of these renormalized low-energy constants)
\eq
X(\mu )= X(\mu ') + {{\Gamma_X}\over{(4\pi )^2}}\,\cdot\ln ({\mu '}/\mu )\ .
\lbl{scale}
\en 
At order ${\cal O}({\rm p}^4)$, the low-energy constants of 
${\tilde{\cal L}}^{(2)}$ and ${\tilde{\cal L}}^{(3)}$ also need to be
renormalized. The corresponding counterterms, however, are of order
${\cal O}(B_0^2)$ and ${\cal O}(B_0)$, respectively, and they are gathered in 
the
three last terms of Eq.  \rf{L4}:  in G$\chi$PT, renormalization
proceeds order by order in the expansion in powers of
$B_0$ \cite{SSF,KS}. Alternatively, one may think of Eqs.  \rf{L2}  and  
\rf{L3} as
standing for the combinations  ${\tilde{\cal L}}^{(2)} + B_0^2{\cal
L}'_{(0,2)}$ and ${\tilde{\cal L}}^{(3)} + B_0{\cal L}'_{(2,1)} +
B_0{\cal L}'_{(0,3)}$, respectively, with the corresponding low-energy
constants representing the renormalized quantities.  The full list of
$\beta$-function coefficients $\Gamma_X$ are tabulated below.

\begin{table}
\begin{center}
\begin{tabular}{|c|c|} \hline
                  $X$   & $F_0^2\cdot\Gamma_X$ \\ \hline
$A_0$   & $5B_0^2/3$\\ 
$Z_0^S$ & $11B_0^2/9$\\ 
$Z_0^P$ & $0$ \\ \hline 
$H_0$   & $10B_0^2/3$ \\ \hline
\end{tabular}
\vskip 0.5 true cm
\caption{Subtraction scale dependences of the low energy
constants of ${\tilde{\cal L}}^{(2)}$.}
\end{center}
\end{table}

\begin{table}
\begin{center}
\begin{tabular}{|c|c|} \hline
                  $X$   & $F_0^2\cdot\Gamma_X$ \\ \hline
$\xi$          & $3B_0$\\ 
${\tilde\xi}$  & $B_0 $\\ \hline
$\rho_1$   & $4B_0\left( \frac{1}{6}A_0 + Z_0^S + Z_0^P \right)$ \\
$\rho_2$   & $4B_0\left( \frac{1}{6}A_0 - Z_0^S + 3Z_0^P \right)$\\
$\rho_3$   & $4B_0\left( {1\over 2}A_0 - {2\over 3}Z_0^S + 
                        {5\over 6}Z_0^P \right)$ \\
$\rho_4$   & $4B_0\left( {11\over 9}A_0 + {5\over 6}Z_0^S -
                        {2\over 3}Z_0^P \right)$ \\
$\rho_5$   & $4B_0\left( A_0 + {5\over 3}Z_0^S - 
                        {4\over 3}Z_0^P \right)$ \\
$\rho_6$   & $4B_0\left( {1\over 9}Z_0^S + {11\over 18}Z_0^P \right)$ \\
$\rho_7$   & $4B_0\left( {11\over 18}Z_0^S + {1\over 9}Z_0^P \right)$ \\
                                                                       \hline
\end{tabular}
\vskip 0.5 true cm
\caption{Subtraction scale dependences of the low energy
constants of ${\tilde{\cal L}}^{(3)}$.}
\end{center}
\end{table}

\begin{table}
\begin{center}
\begin{tabular}{|c|c|} \hline
                  $X$   & $\Gamma_X$ \\ \hline
$L_1$    & ${3\over{32}}$\\ 
$L_2$    & ${3\over{16}}$\\ 
$L_3$    & $0$ \\ 
$L_9$    & ${1\over{4}}$ \\ 
$L_{10}$ & $-{1\over{4}}$ \\ \hline
$H_1$    & $-{1\over{8}}$ \\ \hline
\end{tabular}
\vskip 0.5 true cm
\caption{Subtraction scale dependences of the low energy
constants of ${\cal L}_{(4,0)}$.}
\end{center}
\end{table}

\begin{table}
\begin{center}
\begin{tabular}{|c|c|} \hline
                  $X$   & $F_0^2\cdot\Gamma_X$ \\ \hline
$A_1$      & $-6(Z_0^S - Z_0^P)$\\ 
$A_2$      & $0$\\ 
$A_3$      & $0$ \\ 
$A_4$      & $2(Z_0^S - Z_0^P)$ \\ \hline
$B_1$      & $(3A_0 - 2Z_0^S - 2Z_0^P)$\\ 
$B_2$      & $4(Z_0^S + Z_0^P)$\\ 
$B_3$      & $0$ \\ 
$B_4$      & $(A_0 + Z_0^S + Z_0^P)$ \\ \hline
$C_1^S$    & $(A_0 - 2Z_0^P)$\\ 
$C_2^S$    & $0$\\ 
$C_3^S$    & $0$ \\ 
$C_1^P$    & $(A_0 - 2Z_0^S)$\\ 
$C_2^P$    & $0$\\ 
$C_3^P$    & $0$ \\ \hline 
$D^S$      & $(A_0 + 7Z_0^S)$\\
$D^P$      & $(A_0 + 7Z_0^P)$\\ \hline
$H_2$      & $0$\\ \hline
\end{tabular}
\vskip 0.5 true cm
\caption{Subtraction scale dependences of the low energy
constants of ${\cal L}_{(2,2)}$.}
\end{center}
\end{table}

\begin{table}
\begin{center}
\begin{tabular}{|c|c|} \hline
                  $X$   & $F_0^2\cdot\Gamma_X$ \\ \hline
$E_1$      & $-{7\over 3}A_0^2 + 8A_0(Z_0^S + Z_0^P) + 
                            {34\over 3}(Z_0^S + Z_0^P)^2$\\ 
$E_2$      & $-16A_0(Z_0^S - Z_0^P) + 24(Z_0^S)^2 - 24(Z_0^P)^2$\\ 
$E_3$      & $-{14\over 3}A_0^2 + 8A_0(Z_0^S + Z_0^P) +
               {8\over 9}\left[ 35(Z_0^S)^2 + 35(Z_0^P)^2 
                             - 2Z_0^S Z_0^P\right]$ \\ \hline 
$F_1^S$    & ${26\over 9}A_0^2 - {25\over 9}( Z_0^S + Z_0^P )^2$\\
$F_2^S$    & $2A_0^2 + {2\over 3}A_0 Z_0^S - {16\over 3}A_0 Z_0^P -
             {4\over 9}\left[ 17(Z_0^S)^2 + 8(Z_0^P)^2 + 
                                        25 Z_0^S Z_0^P\right]$\\
$F_3^S$    & $2A_0^2 + {2\over 3}A_0 Z_0^S - {16\over 3}A_0 Z_0^P -
             {4\over 9}\left[ 71(Z_0^S)^2 + 8(Z_0^P)^2 -
                                        29 Z_0^S Z_0^P\right]$\\
$F_4^S$    & $-20 (Z_0^S)^2 + 20 (Z_0^P)^2$\\ \hline
$F_1^P$    & $A_0^2 - {8\over 9}( Z_0^S + Z_0^P )^2$\\
$F_2^P$    & $2A_0^2 - {16\over 3}A_0 Z_0^S + {2\over 3}A_0 Z_0^P -
             {4\over 9}\left[ 8(Z_0^S)^2 + 17(Z_0^P)^2 + 
                                        25 Z_0^S Z_0^P\right]$\\
$F_3^P$    & $-2A_0^2 + {16\over 3}A_0 Z_0^S - {2\over 3}A_0 Z_0^P +
             {4\over 9}\left[ 8(Z_0^S)^2 + 71(Z_0^P)^2 -
                                        29 Z_0^S Z_0^P\right]$\\ \hline
$F_5^{SS}$   & ${44\over 9}A_0 Z_0^S + {8\over 9}A_0 Z_0^P +
                {1\over 9}\left[85(Z_0^S)^2 + 4(Z_0^P)^2 +
                                       26Z_0^S Z_0^P \right]$\\
$F_6^{SS}$   & $4A_0 Z_0^S +{2\over 9}\left[77(Z_0^S)^2 -
                  4(Z_0^P)^2 - 26Z_0^S Z_0^P \right]$\\
$F_5^{SP}$   & ${8\over 9}A_0 Z_0^S + {44\over 9}A_0 Z_0^P +
                {1\over 9}\left[4(Z_0^S)^2 + 85(Z_0^P)^2 +
                                       26Z_0^S Z_0^P \right]$\\
$F_6^{SP}$   & $-4A_0 Z_0^P +{2\over 9}\left[4(Z_0^S)^2 -
                  77(Z_0^P)^2 + 26Z_0^S Z_0^P \right]$\\
$F_7^{SP}$   & $2A_0(Z_0^S + Z_0^P) - {2\over 9}\left[4(Z_0^S)^2
                          + 4(Z_0^P)^2 - 55Z_0^S Z_0^P \right]$\\ \hline
$H_3$      & $ 8A_0(Z_0^S + Z_0^P) + {4\over 9}\left[35(Z_0^S)^2
                          + 35(Z_0^P)^2 - 2Z_0^S Z_0^P \right]$\\ 
$H_4$      & $ 4A_0^2 - {4\over 9}\left[17(Z_0^S)^2
                          + 17(Z_0^P)^2 - 2Z_0^S Z_0^P \right]$\\ \hline
\end{tabular}
\vskip 0.5 true cm
\caption{Subtraction scale dependences of the low energy
constants of ${\cal L}_{(0,4)}$.}
\end{center}
\end{table}

\pagebreak
\clearpage

\setcounter{equation}{0}
\addtocounter{zahler}{1}
\section{Masses and Decay Constants}

For the reader's convenience and for later reference, we provide, in this 
Appendix, the expressions of the pseudoscalar decay constants and masses 
at order 
${\cal O}({\rm p}^4)$ \footnote{We also take this opportunity to correct a 
misprint in an earlier published expression \cite{KMSF} 
of the decay constant $F_K$.}.
Introducing the notation ($P=\pi,K,\eta$)
\be
\mu_{P}\,=\,\frac{M_P^2}{32\pi^2F_\pi^2}\,\ln\frac{M_P^2}{\mu^2}\;,
\en
where $\mu$ is the renormalisation scale, we obtain the following formulae for 
the decay constants (in the limit $m_u=m_d$),

\be
F_{\pi}^2 = F_0^2\left[ 1 + 2{\widehat m}\delta_{F,\pi}^{(2,1)} +
                            2{\widehat m}^2\delta_{F,\pi}^{(2,2)}
                          - 4\mu_{\pi} - 2\mu_K\right]\ ,
\en
with
\bea
\delta_{F,\pi}^{(2,1)}&=&\xi + (2+r){\tilde\xi}\nonumber\\
\quad\\
\delta_{F,\pi}^{(2,2)}&=&A_1 + {1\over 2}A_2 + 2A_3 + B_1 - B_2 \nonumber\\
        &&+{1\over 2}(2+r^2)(A_4 + 2B_4) + 2(2+r)D^S\nonumber
\ena

\be
F_{K}^2 = F_0^2\left[ 1 + 2{\widehat m}\delta_{F,K}^{(2,1)} +
                            2{\widehat m}^2\delta_{F,K}^{(2,2)}
               - {3\over 2}\mu_{\pi} - 3\mu_K -{3\over 2}\mu_{\eta}\right]\ ,
\en
with
\bea
\delta_{F,K}^{(2,1)}&=&{1\over 2}(1+r)\xi + (2+r){\tilde\xi}\nonumber\\
\quad\\
\delta_{F,K}^{(2,2)}&=&{1\over 2}(1+r^2)(A_1 + A_3 + B_1) + {r\over 2}
             ( A_2 + 2A_3 - 2B_2)\nonumber\\
        &&+{1\over 2}(2+r^2)(A_4 + 2B_4) + (1+r)(2+r)D^S\nonumber
\ena

\be
F_{\eta}^2 = F_0^2\left[ 1 + 2{\widehat m}\delta_{F,\eta}^{(2,1)} +
                            2{\widehat m}^2\delta_{F,\eta}^{(2,2)}
                          - 6\mu_K\right]\ ,
\en
with
\bea
\delta_{F,\eta}^{(2,1)}&=&{1\over 3}(1+2r)\xi + (2+r){\tilde\xi}\nonumber\\
\quad\\
\delta_{F,\eta}^{(2,2)}&=&{1\over 3}(1+2r^2)
                         (A_1 + {1\over 2}A_2 + 2A_3 + B_1 - B_2)\nonumber\\
   &&+{1\over 2}(2+r^2)(A_4 + 2B_4) + {2\over 3}(1+2r)(2+r)D^S\nonumber\\
        &&-{4\over 3}(r-1)^2(C_1^P + C_2^P)\nonumber
\ena

For the masses, we obtain:
\bea
F_{\pi}^2M_{\pi}^2 &=& F_0^2 \bigg\{ 2{\widehat m}B_0 +4{\widehat m}^2 A_0 +
                               4{\widehat m}^2 (2+r)Z_0^S\nonumber\\
&&\qquad + 2{\widehat m}^3 \delta_{M,\pi}^{(0,3)} + 2{\widehat m}^4
\delta_{M,\pi}^{(0,4)} + 4{\widehat m}^2 A_3 M_{\pi}^2\nonumber\\
&&\qquad  - \, \mu_{\pi} \, [3\,M_{\pi}^2 + 5 \, M_{\pi}^2 \,
\hat{\epsilon}(r) ]  \\
&&\qquad  - 2 \, \mu_K     \, [M_{\pi}^2 + (1+r)\,M_{\pi}^2 \,
\hat{\epsilon}(r) ] \nonumber \\
&&\qquad  - {1 \over 3}\,\mu_{\eta} \bigg[M_{\pi}^2 + (1+2r)\,M_{\pi}^2 \,
\hat{\epsilon}(r)  - 2\,{\Delta_{GMO} \over r-1}\bigg]
\bigg\} 
\nonumber 
\ena

\bea
F_{K}^2M_{K}^2 &=& F_0^2 \bigg\{ (1+r){\widehat m}B_0 +
                            (1+r)^2{\widehat m}^2 A_0 +
                            2(1+r)(2+r){\widehat m}^2 Z_0^S\nonumber\\
&&\qquad + (1+r){\widehat m}^3 \delta_{M,K}^{(0,3)} + (1+r){\widehat m}^4
\delta_{M,K}^{(0,4)} + (1+r)^2{\widehat m}^2 A_3 M_{K}^2\nonumber\\
&&\qquad - {3 \over 2} \, \mu_{\pi} \, [M_K^2 + M_{\pi}^2 \,
\hat{\epsilon}(r) ] \\ 
&&\qquad - 3 \,           \mu_K \,   \,  [M_K^2 + {1 \over 2}
\,(1+r)^2\,M_{\pi}^2 \, \hat{\epsilon}(r) ] \nonumber \\ 
&&\qquad - {1 \over 6}\,  \mu_{\eta}  \, \bigg[ 5\,M_K^2 +
(r+1)\,(1+2r)\,M_{\pi}^2 \, \hat{\epsilon}(r)  + { r+1
\over r-1}\,\Delta_{GMO} \bigg] \bigg\}\nonumber  
\ena

\bea
F_{\eta}^2M_{\eta}^2 &=& F_0^2 \bigg\{ {2\over 3}(1+2r){\widehat m}B_0 +
                    {4\over 3}(1+2r^2){\widehat m}^2 A_0 +
                    {4\over 3}(1+2r)(2+r){\widehat m}^2 Z_0^S + 
                    {8\over 3}(1-r)^2{\widehat m}^2 Z_0^P \nonumber\\
&&\qquad + 2{\widehat m}^3 \delta_{M,\eta}^{(0,3)} + 2{\widehat m}^4
\delta_{M,\eta}^{(0,4)} + 
{4\over 3}(1+2r^2){\widehat m}^2 A_3 M_{\eta}^2 -
{8\over 3}(1-r)^2{\widehat m}^2 C_2^P M_{\eta}^2\nonumber\\
&&\qquad - \,\mu_{\pi} \, \bigg[ M_{\pi}^2 + (2r+1) \, M_{\pi}^2 \,
\hat{\epsilon}(r)  - 2 \, {\Delta_{GMO} \over r-1} \bigg]  \\ 
&&\qquad - {2 \over 3} \,\mu_K \, \bigg[ 8 M_K^2 - 3 M_{\pi}^2 + (r+1)
(2r+1) \, M_{\pi}^2 \, \hat{\epsilon}(r)  + 2\,{2r-1 
\over r-1}\, \Delta_{GMO} \bigg] \nonumber  \\  
&&\qquad - {1 \over 9} \,\mu_{\eta} \, \bigg[16 M_K^2 - 7 M_{\pi}^2 + 3\,
(2r+1)^2 \,M_{\pi}^2 \, \hat{\epsilon}(r)  + 4\,{4r-1 \over r-1}\,
\Delta_{GMO}  \bigg] \bigg\} \nonumber \ ,
\ena
with
\bea
\delta_{M,\pi}^{(0,3)}&=&{9\over 2}\rho_1 + {1\over 2}\rho_2 +
                         (10+4r+r^2)\rho_4 + {1\over 2}(2+r^2)\rho_5 +
                         6(2+r)^2\rho_7\nonumber\\
&&\qquad\\ 
\delta_{M,K}^{(0,3)}&=&{3\over 2}(1+r+r^2)\rho_1 + {1\over 2}(1-r+r^2)\rho_2 +
                         3(2+2r+r^2)\rho_4 + {1\over 2}(2+r^2)\rho_5 +
                         6(2+r)^2\rho_7\nonumber\\
&&\qquad\\ 
\delta_{M,\eta}^{(0,3)}&=&{3\over 2}(1+2r^3)\rho_1 + {1\over 6}(1+2r^3)\rho_2 +
                          {8\over 3}(1+r)(1-r)^2 \rho_3 + \nonumber\\
&&\qquad                  {1\over 3}(10+8r+17r^2+10r^3)\rho_4 + 
                          {1\over 6}(1+2r)(2+r^2)\rho_5 + \nonumber\\ 
&&\qquad                  {8\over 3}(1-r)^2 (2+r)\rho_6 +
                          2(1+2r)(2+r)^2\rho_7\ ,
\ena
while the ${\cal O}({\rm p}^4)$ contributions are
\bea
\delta_{M,\pi}^{(0,4)}&=& 8E_1 + 2E_2 + 4 (1+r)(2+r^2)F_1^S + 
                          (20+9r+r^3)F_2^S \nonumber\\
&&\qquad + (4+r+r^3)F_3^S 
 + 2(2+r^2)F_4^S + 4(2+r)(6+2r+r^2)F_5^{SS}\\
&&\qquad +
2(2+r)(2+r^2)F_6^{SS}\nonumber\ ,\\ \nonumber \\
\delta_{M,K}^{(0,4)}&=& 2(1+r)(1+r^2)E_1 + (1+r^3)E_2 +
       {1\over 2} (1+r)(1-r)^2E_3\nonumber \\
&&\qquad + 4(1-r)(2+r^2)F_1^S + (8+9r+9r^2+4r^3)F_2^S \nonumber\\
&&\qquad + (4-r+r^2+2r^3)F_3^S 
 + (1+r)(2+r^2) F_4^S \\
&&\qquad + 4(2+r)(4+3r+2r^2)F_5^{SS} +
         2(2+r)(2+r^2)F_6^{SS}\nonumber\ ,\\ \nonumber \\
\delta_{M,\eta}^{(0,4)}&=& {8\over 3}(1+2r^4)E_1 + {2\over 3}(1+2r^4)E_2
\nonumber\\
&&\qquad +    {8\over 3}(1+2r^2)(2+r^2)F_1^S + {1\over
3}(20+13r+37r^3+20r^4)F_2^S  \nonumber\\ 
&&\qquad +{1\over 3}(4+5r+5r^3+4r^4)F_3^S
                  + {2\over 3}(1+2r^2)(2+r^2)F_4^S\nonumber\\
&&\qquad + {16\over 3}(1+r)^2(1-r)^2F_1^P +4(1-r)^2(1+r+r^2)F_2^P \nonumber\\
&&\qquad +
                    {4\over 3}(1-r)^2(1+r+r^2)F_3^P
+ 4(2+r)(2+2r+3r^2+2r^3)F_5^{SS} \nonumber\\
&&\qquad
         + {2\over 3}(2+r)(2+r^2)(1+2r)F_6^{SS}
 + {8\over 3}(1-r)^2(2+r^2)F_5^{SP}\\
&&\qquad         + {4\over 3}(1-r)^2(2+r^2)F_6^{SP}
         + {16\over 3}(1-r)^2(1+r)(2+r)F_7^{SP}\ .\nonumber
\ena

\pagebreak

\end{document}